# An Efficient Algorithm for Topological Characterisation of Worm-Like and Branched Micelle Structures from Simulations

Breanndan O Conchuir,*,† Kirk Gardner,‡ Kirk E. Jordan,¶ David J. Bray,§
Richard L. Anderson,§ Michael A. Johnston,∥ William C. Swope,⊥ Alex Harrison,†
Donald R. Sheehy,# and Thomas J. Peters*,‡

†IBM Research U.K., The Hartree Centre, Daresbury WA4 4AD, United Kingdom

‡Department of Computer Science & Engineering, University of Connecticut, Storrs, CT
06269, United States

¶IBM T.J. Watson Research, Cambridge, Massachusetts, United States

§The Hartree Centre, STFC Daresbury Laboratory, Warrington, WA4 4AD, United
Kingdom

∥IBM Research Ireland, Dublin, Ireland

⊥IBM Almaden Research Center, San Jose, California, United States

#Department of Computer Science, North Carolina State University, Raleigh, NC 27695,
United States

E-mail: breanndan.conchuir@ibm.co.uk; tpeters@engr.uconn.edu

## Abstract

Many surfactant-based formulations are utilised in industry as they produce desirable visco-elastic properties at low-concentrations. These properties are due to the presence of worm-like micelles (WLM) and, as a result, understanding the processes that lead to WLM formation is of significant interest. Various experimental techniques have been applied with some success to this problem but can encounter issues probing key microscopic characteristics or the specific regimes of interest. The complementary use of computer simulations could provide an alternate route to accessing their structural and dynamic behaviour. However, few computational methods exist for measuring key characteristics of WLMs formed in particle simulations. Further, their mathematical formulation are challenged by WLMs with sharp curvature profiles or density fluctuations along the backbone. Here we present a new topological algorithm for identifying and characterising WLMs micelles in particle simulations which has desirable mathematical properties that address short-comings in previous techniques. We apply the algorithm to the case of Sodium dodecyl sulfate (SDS) micelles to demonstrate how it can be used to construct a comprehensive topological characterisation of the observed structures.

# 1 Introduction

Micelles play an important role in many processes and products. For example, micelles can be used to control drug release,[1–7] are key in household cleaning products,[8,9] and can act as friction modifiers in vehicle engines.[10,11] They display a wide variety of fluctuating behaviours and morphologies depending upon the surfactant from which they are assembled and the environment in which they reside. The different structures formed by micelles give rise to multiple possible behaviours including interesting optical and rheological properties. For example, rod- and worm-like micelles (WLMs) are associated with a significant increase in overall viscosity of a liquid when compared to spherical micelles. WLMs are long flexible cylinders with diameters (for typical surfactants molecules) of approx 3-5 nm and persistence lengths of approx. 20 nm below which they should be considered as rigid rods. These micelles are able to grow to a length of several micrometers at relatively low surfactant concentrations resulting in the occurrence of entanglements and ultimately enhanced visco-elasticity.[12–15]

Under certain conditions WLMs may branch between each other to form interconnected networks. For example, electrostatic screening by counter-ions is known to promote WLM branch formation.[16–18] Whilst long entangled micelles lead to increased zero-shear viscosity, branching and network formation is generally believed to lower it.[19,20] Thermal fluctuations can cause micelle branches to slide along the main micelle backbone chain thereby providing an accessible mode of stress relaxation.[19,21] It is believed this in turn leads to a reduction in viscosity when compared to an entangled matrix of WLMs.[17,21–24]

The formation of WLM structures is dependent upon the underlying structural characteristics of the surfactants molecules from which they are constructed. Surfactants that form WLMs preferentially pack into cylindrical shapes. Additives can significantly alter this preference through, for example, hydrophobic binding and electrostatic screening, leading to changes in WLM structure and rheological properties.

Exploring the behaviour of WLMs can be challenging using experimental methods. Cryo-TEM gives the researcher a window into the micelle structure but this approach can be difficult because of the high viscosities of concentrated surfactant solutions. Pulsed gradient spin-echo nuclear magnetic resonance (PGSE-NMR)[25] and multiple scattering approaches[26,27] can supply information on the branching behaviour of micelles. However, branching is most often inferred from rheological observations (i.e., a decrease in the zero shear viscosity).

Computer simulations offer a complementary approach to existing experimental techniques for studying WLMs and are beginning to shine some light onto the structural and dynamic behaviour of WLMs. Self-assembly of surfactant molecules into micelles is difficult to simulate with all-atom methods, such as Molecular Dynamics (MD), due to the high computational cost associated with obtaining appropriate length and time-scales. Consequently only a few micelles can currently be studied with these techniques. As such, coarse-grained (CG) approaches are often applied in the study of micelles, such as the MARTINI force-field.[9,16,18,28–40] Dissipative Particle Dynamics (DPD) is another CG approach which is becoming increasingly popular for the simulation of micelle behaviour as it allows signif-icantly longer length and time-scales to be accessed at reduced computational cost (albeit with a potential loss in accuracy).[41–43] DPD employs very soft conservative potentials which facilitates the use of large time-steps. Further details of the DPD method can be found in the literature.[41–44] A recent review on simulating surfactants and micelle systems (including WLMs) using a variety of particle-based simulation techniques has been presented by Taddese et al.[45] In addition, a review of the application of computer simulation to the study of WLMs can be found in reference.[46]

A number of rudimentary analytical approaches have been used to characterise micelles including aggregation numbers and size distributions, radius of gyration, asphericity[47] which are often adequate for small micelles with simple topological structures but less appropriate for larger and more topologically complex structures such as WLMs. Here quantification of branch points, micelle backbone length and end-cap formation are more important but far less methods exist in the literature for capturing these properties. A notable exception is the work of Dhakal and Sureshkumar who have employed a contour mesh measurement approach to calculate a micelle backbone length scale and identify micelle branch points and end-caps.[16] This model was later successfully employed to examine the stress relaxation modes of branched micelles.[21] Nevertheless, their contour length measurement algorithm is mathematically limited to piecewise linear backbone length metrics where each constituent line is of equal length. A more comprehensive topological framework is required to accurately represent the geometric complexity of micelles with sharp curvature profiles or density fluctuations along the backbone. In particular, these latter considerations cannot be overlooked when one wishes to simulate multi-component industrial formulations where surfactant polydispersity, and the presence of co-surfactants and modifiers potentially require additional scrutiny.

The remainder of this article is arranged as follows. First, we give an overview of the topological algorithm applied to characterise the salient geometric length scales of WLMs and branches. Next, we detail the DPD simulation protocol and SDS micelle system test case considered in this study. We then present our observations for the SDS test case. The paper concludes with a discussion on the developed topological algorithm including how some of the computed metrics can be utilised as input parameters to macroscopic micelle viscoelasticity models.

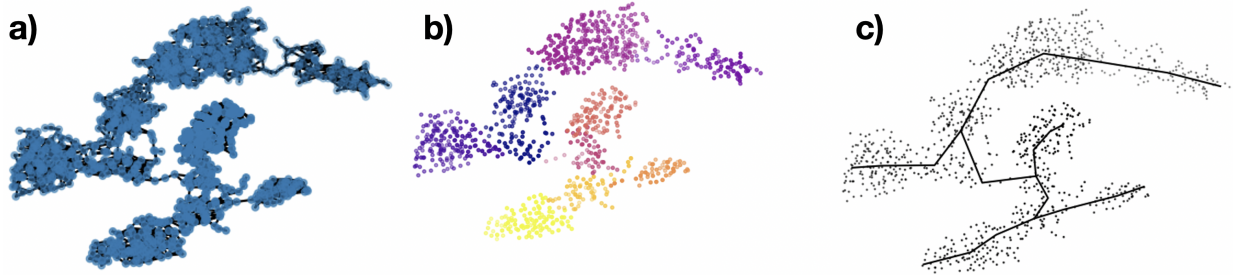

Figure 1: The key stages in the application of the topological algorithm. a) Micelle point cloud, b) structural segmentation, and c) geometric backbone network.

## 2 Characterisation of Micelle Topology

### 2.1 Algorithmic Design

The framework behind the topological characterisation algorithm introduced in this article is outlined in Fig. 1. First, individual micelles are identified using trajectory data from DPD simulations (See Section 3.1). Each micelle can be represented in the form of a point cloud (and neighbourhood graph) which is the set of 3D spatial coordinates of each molecular bead in the cluster (Fig. 1a). Next, the point cloud structure undergoes an operation known as Fiedler vector segmentation which partitions the structure into a number of distinct regions with corresponding centre points (Fig. 1b). This exercise is repeated until an optimal spectral clustering is achieved which captures important features such as branch points and loops without overfitting (Fig. 2). Finally, these points are joined together to produce a piecewise linear (PL) skeleton of the micelle (Fig. 1c) composed of a set of consecutive linear segments of varying individual lengths. This key step allows us to calculate a series of structural micelle characteristics such as the micelle backbone length, the number of branch points and the Euclidean distance between end-caps.

A description of the theory underpinning the mathematical framework is presented in the Supporting Information.

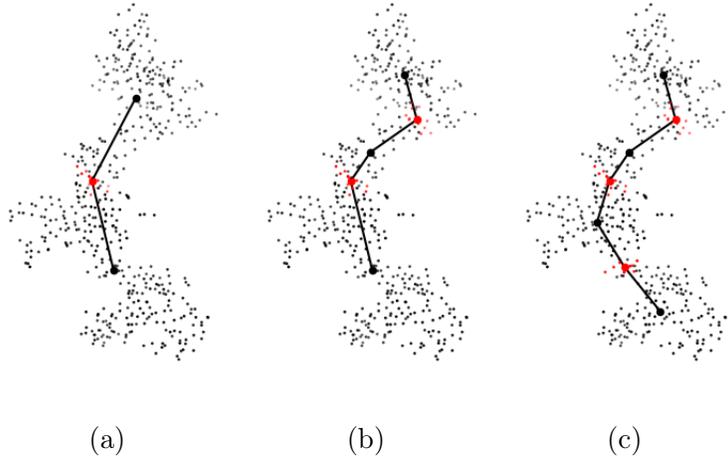

(a)          (b)          (c)

Figure 2: Recursive spectral segmentation of a micelle point cloud into four subsets corresponding to the black vertices. Red vertices indicate links between these subsets.

## 2.2 Spectral Splits

The spectral gap, denoted here by $\lambda$, is equal to the first eigenvalue of the discrete Laplacian.[48] The general protocol is to recursively split sets at their spectral gaps (with further information outlined in the Supporting Information). This process works well for large WLMs; however, overfitting from unnecessary cuts leads to excess length and inaccurate branch points in smaller micelles without clear WLM structure. To avoid this, we introduce a bound $\lambda_{max}$ on the spectral gap $\lambda$ of each potential split which ensures a sufficiently sparse cut. We note that recursion introduces new eigenvalues for each segment generated and new spectral gaps. On these data sets, we have observed that the spectral gaps increase during recursion.

There are two fundamental properties of a good split that can be revealed by the spectral gap:

a. **Size of subsets:** The resulting subsets are sufficiently large and contain approximately

the same number of vertices.

b. **Number of connecting edges:** The number of edges extending from one subset to the other is small.

Figure 3 depicts splits which violate these properties, with the two subgraphs resulting from the split in red and blue, and edges between vertices surrounding the split in black.

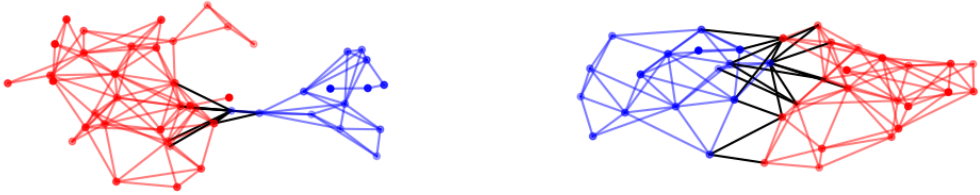

(a) One subset (blue) is too small ($\lambda = 0.0426$).

(b) Too many connecting edges. ($\lambda = 0.1085$).

Figure 3: Cuts where the spectral gap reveals violations of the properties of a "good" split.

Empirically, we have discovered that terminating the recursion when any newly computed spectral gap value exceeds $\lambda_{max} = 0.015$ has been a good stopping criterion over the data tested. Since this was determined empirically, there is no claim that the value of 0.015 is optimal.

## 2.3 Implementation

### 2.3.1 Segmentation

The recursive procedure SEGMENT is detailed in Algorithm 1. For each recursive step the spectral gap $\lambda$ is used as a stopping condition so that recursion can only continue when $\lambda > \lambda_{max}$. To avoid overfitting we impose a soft maximum recursion depth $k_{max}$. To avoid underfitting, we set a lower bound $\lambda_{min}$ on the spectral gap so that $k_{max}$ only applies when $\lambda_{min} \leq \lambda \leq \lambda_{max}$. It can be shown that the gap of each of the resulting subgraphs must be greater than $\lambda$ and continued partitioning will converge to $\lambda_{min}$.

**Algorithm 1** Recursive segmentation.

**Require:** $k_{max}, \lambda_{min}, \lambda_{max}$ ▷ Global parameters.
  **function** SEGMENT($G = (V, E), \ k = 0$) ▷ Input graph $G$ and depth counter $k$.
    $L \leftarrow$ NORMALIZEDLAPLACIAN($G$) ▷ Compute the normalized Laplacian of $G$.
    $\{(\lambda_i, w_i)\}_{\lambda_i \leq \lambda_{i+1}} \leftarrow$ EIGENDECOMP($L$) ▷ Eigenvalues $\lambda_i$ and vectors $w_i$ sorted by $\lambda_i$.
    $\lambda, f \leftarrow \lambda_1, w_1$ ▷ Ensure $0 = \lambda_0 < \lambda_1$.
    **if** $\lambda < \lambda_{max}$ **then**
      $G_1, G_2 \leftarrow$ SPLIT($G, f$) ▷ $G_1, G_2$, and link stored in structure.
      **if** $k < k_{max}$ **or** $\lambda < \lambda_{min}$ **then** ▷ Continue if gap is still small.
        SEGMENT($G_1, k+1$) ▷ Recurse on $G_1$.
        SEGMENT($G_2, k+1$) ▷ Recurse on $G_2$.
      **end if**
    **end if**
  **end function**

The SPLIT procedure, detailed in Algorithm 2, takes a graph $G$ and partitions its vertices by the vector $f$. For each split the resulting graph partition has vertices corresponding to subgraphs $G_1$ and $G_2$ which are stored in a tree structure $T$ with `children`($G$) = $\{G_1, G_2\}$. If $G$ is connected, a single split will result in one edge added to the graph partition which corresponds to a link subgraph $G_l$ that associates its children. This association captures the vertices in the graph partition which are split by $G_l$ and will be denoted by `nbr`($G_l$) so that `nbr`($G_l$) = $\{G_1, G_2\}$ initially. The asymptotic complexity is bounded above, quadratically,[49] by the eigendecomposition of a graph Laplacian, but implementation optimizations greatly improve the amortized running time in practice.

Although each step of the segmentation resembles a graph partition the resulting structure is not, strictly speaking, a graph partition. This is due to the fact that when child subgraphs $G_1$ and $G_2$ are split further, `nbr`($G_l$) is updated to include the children of $G_1$ and $G_2$ containing vertices in $G_l$. Letting `leaf`($T$) = $\{H_1, \ldots, H_m\}$ denote the set of subgraphs corresponding to leaf nodes and `links`($T$) = $\{L_1, \ldots, L_k\}$ denote the collection of links at any point in the segmentation we have

$$\texttt{nbr}(L_j) = \{H_i \in T \mid H \cap L_j \neq \emptyset\},$$

for all $L_j \in \texttt{links}(T)$, where the intersection is taken as the intersection of vertex sets. Note that a link node may associate more than two leaf nodes, indicating a branch point in the skeleton. An additional step merges intersecting links under certain conditions. The resulting structure is therefore not a graph partition as we treat the links as vertices of the skeleton which may have degree greater than two.

---

**Algorithm 2** The split procedure.

---

**Require:** Tree $T$ with subgraphs as nodes.
    Stores parent / child relationships and links between leaf nodes.
    **function** SPLIT($G = (V, E)$, $f$)                               ▷ Input graph $G$ and Fiedler vector $f$
        Split $G$ into subgraphs $G_1, G_2$ and $G_l$ by $f$.
        Set $\texttt{children}(G) = \{G_1, G_2\}$.
        Set $\texttt{nbr}(G_l) = \{G_1, G_2\}$ and add $G_l$ to $\texttt{links}(T)$.
        **for** $L \in \texttt{links}(T)$ such that $G \in \texttt{nbr}(L)$ **do**
            Remove $G$ from $\texttt{nbr}(L)$.
            **for** $H \in \texttt{children}(G)$ such that $H \cap L \neq \emptyset$ **do**
                Add $H$ to $\texttt{nbr}(L)$.
            **end for**
            **if** $G_l \cap L \neq \emptyset$ **then**
                Merge links $G_l$ and $L$.
            **end if**
        **end for**
        **return** $G_1, G_2$
    **end function**

---

When the segmentation procedure terminates the leaf nodes of the resulting tree structure correspond to a collection of subgraphs with vertex sets that partition the input vertices. We will refer to these leaf subgraphs as $\texttt{leaf}(T) = \{H_1, \ldots, H_m\}$ and the collection of link subgraphs as $\texttt{links}(T) = \{L_1, \ldots, L_k\}$.

### 2.3.2 Skeletonization

Recall that each vertex $v_i$ in a micelle graph $G = (V, E)$ is associated with a point $p_i \in \mathbb{R}^3$ that gives the location of the corresponding DPD bead in the micelle. Therefore, the vertex set of a subgraph $G' = (V', E')$ of $G$ can be associated with a collection of points in $\mathbb{R}^3$,

denoted
$$\overline{G'} = \{p_i \in \mathbb{R}^3 \mid v_i \in V'\}.$$

Let $\phi$ be the centroid function, which maps subsets of $\mathbb{R}^3$ to a single point, by

$$\phi(P) = \frac{1}{|P|} \sum_{p \in P} p$$

where $|P|$ denotes the cardinality of $P$.

For a micelle graph $G = (V, E)$ the segmentation procedure yields a tree $T$ such that $\texttt{leaf}(T) = \{H_1, \ldots, H_m\}$ is a collection of subgraphs and $\texttt{links}(T) = \{L_1, \ldots, L_k\}$ is a collection of links. Note that the vertex sets of subgraphs in $\texttt{leaf}(T)$ partition $V$ and the edge sets of subgraphs in $\texttt{leaf}(T) \cup \texttt{links}s(T)$ and partition $E$. The *segmentation graph* of $G$ is defined as a graph $S = (V_S, E_S)$ with vertex set

$$V_S = \texttt{links}(T) \cup \texttt{links}(T)$$

and edge set

$$E_S = \{\{L_j, H_i\} \mid L_i \in \texttt{links}(T) \text{ and } H_i \in \texttt{nbr}(L_j)\}.$$

The *skeleton* of a segmentation graph by a function $\phi$ is the graph $\overline{S}^\phi = (\overline{V_S}^\phi, \overline{E_S}^\phi)$ with vertex set

$$\overline{V_S}^\phi = \{\phi(v) \in \mathbb{R}^3 \mid v \in V_S\}$$

and edge set

$$\overline{E_S}^\phi = \{\{\phi(\overline{u}), \phi(\overline{v})\} \mid \{u, v\} \in E_S\}$$

and corresponds to a collection of line segments in $\mathbb{R}^3$ drawn between the centroids of points in the leaves and links of $T$.

### 2.3.3 Estimating the Full Length

We first produce an approximating PL curve that connects the centroids. This curve remains within the interior of the point cloud, so its length is an underestimation. We add additional *cap nodes* to approximate the full length. At each terminal point of the curve, extend the containing segment until it exits the point cloud. The cap node is that exit point. Figure 4 depicts a case in which the cap nodes yield a far more accurate calculation of length.

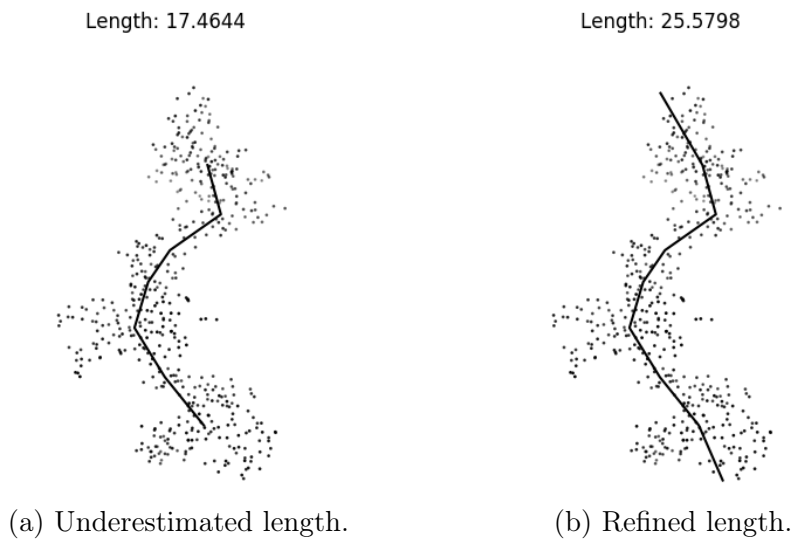

(a) Underestimated length.      (b) Refined length.

Figure 4: The addition of cap nodes has a significant effect on the approximated length.

### 2.3.4 Skeleton Repair

After initial skeletonization, additional segmentation may be required for more complex micelles, in particular for those with branched structure. Fig. 5a depicts a case in which the skeleton does not accurately capture the micelle's structure, resulting in a missed cap node. This is due to a degenerate case in which an end segment has two connections which are too close (Fig. 5b). The final step repairs any degenerate end segments, ensuring that the resulting skeleton accurately reflects both the length and the branching structure of the micelle (Fig. 5c).

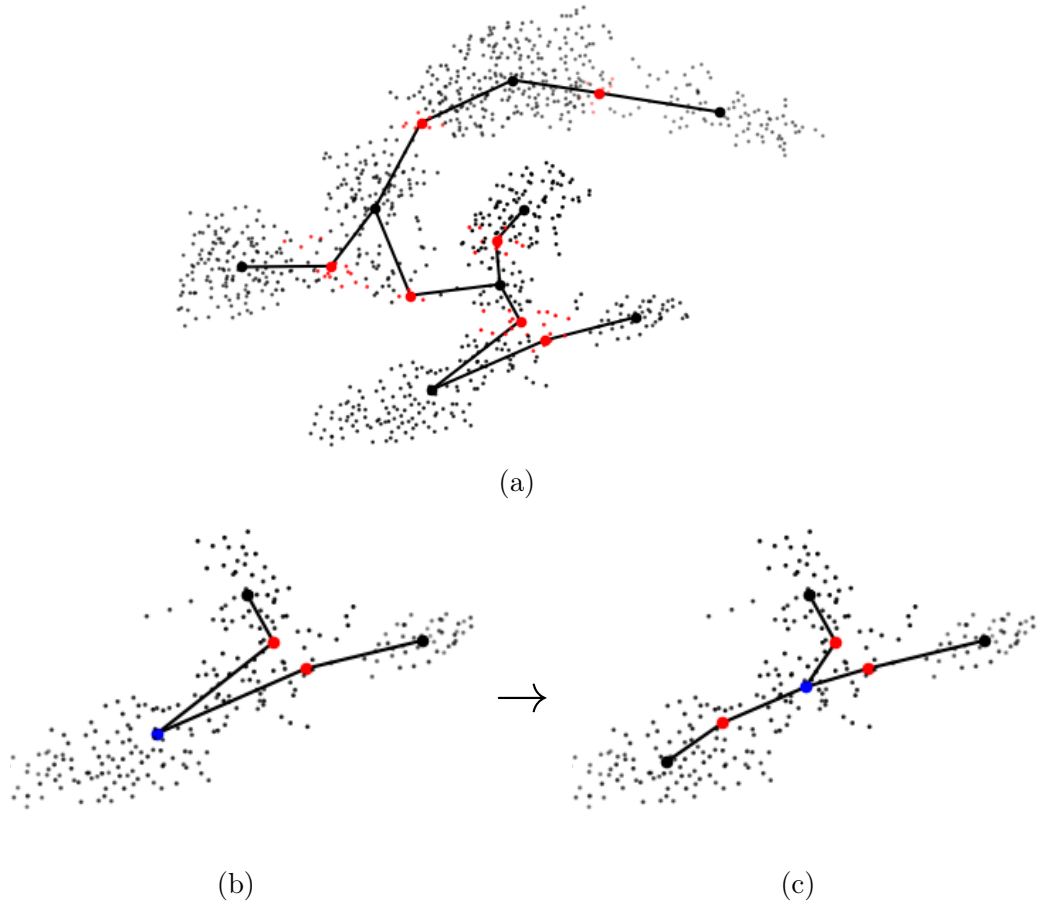

(a)

(b)     (c)

Figure 5: A degenerate segment, corresponding to the vertex in blue, is repaired.

Fig. 5 shows a representative situation, as the blue node has two neighboring nodes, shown in red. The Euclidean distance between the two red nodes is quite small compared to the distance between them, measured along the PL curve. This intuition is captured as a bound on the ratio of the two distances. In particular, we repair a subset of the PL curve when this ratio is less than 0.2. In Fig. 5c the repair is indicated by a new blue node, as a common point for two new subsets of a new PL curve.

### 2.3.5   Estimating a Cross-sectional Radius

After the skeleton has been stabilized, an estimate of a cross-sectional radius for each WLM can be computed. The role of this radius is to approximate a tube of fixed radius, around the skeleton, so that all points of the WLM lie within the tube. For each cluster, the minimum

Euclidean distance from each cluster point to the local linear segment of the skeleton is computed, producing a local radius for each cluster. The final estimate $r_{cs}$ for the cross-sectional radius is the minimum over all these radii as a reliable lower bound.

## 3 Simulation Protocol

To explore the capabilities of the topological characterisation algorithm we performed nine Dissipative Particle Dynamics (DPD) simulations of mixtures of SDS, water and NaCl using the DLMESO simulation package.[50] SDS was chosen as our test case as it is known to readily form WLMs in the presence of salt.[51] Each simulation cell contained 10 w/w% SDS, a fixed amount of NaCl (ranging between 0 - 8 w/w% ), and water. Together, these simulations are analogous to performing a salt curve experiment.[51] An overview of the model used is presented in the Appendix.

Simulations were started from a random configuration and run for a total of 8 million time steps under isothermal-isobaric conditions (NPT) corresponding to room temperature ($K_B T = 1$, 298 K) and at one atmosphere pressure (23.7 in DPD units[42]) with a time step of 0.02. A Langevin piston barostat was employed to control the pressure.[52] The initial simulation box had side lengths of 40 DPD units (22.6 nm following our DPD to real unit mapping). with periodic boundary conditions applied in all three Cartesian directions. An established automated time series equilibration protocol was employed for each observable time series. (See the Supporting Information for further details.)

### 3.1 Micelle Identification and Characterisation

During the simulation, the SDS molecules self-organise into a range of aggregates, transient in nature and of various sizes. At each sampled time frame we used a distance clustering algorithm to determine each aggregate provided by the analytics program UMMAP.[47] An aggregate is defined as the complete set of molecules where each molecule lies no further than 1 DPD unit from another molecule as defined by the distance between beads of two different molecules. Each bead of the aggregate becomes a vertex ($v_i$) of the aggregate and the vector between the pairs of beads that satisfies the distance criteria are referred to as an edge ($e_{ij}$) between the corresponding vertices $v_i$ and $v_j$. A micelle was defined as an aggregate composed of 15 or more constituent surfactant molecules.[53] Smaller aggregates are assigned to be monomers. From this we calculated the following properties: the number-averaged $N_s$ and weight-averaged $N_{ws}$ micelle size, the maximum micelle size $N_{max}$ and the number of micelles $N_m$.

The topology algorithm was then applied to each micelle and from this several properties were calculated which included: cross-sectional radius $r_{cs}$, the number-averaged $l_b$ and weight-averaged $l_{wb}$ micelle backbone lengths, the number-average end-to-end length $l_e$ and number of branched micelles $N_{b,m}$. Here a branched micelle is defined as one with three or more end-caps, while the end-to-end length involves the summation of the Euclidean distance between each pair of end-caps in the micelle. A micelle curvature ratio parameter $\alpha = l_b/l_e$ was also applied as a simple metric to describe the curvature of the micelle worms.[54–56] A full mathematical derivation of how each observable is calculated is provided in the supporting information.

## 4 Test Case: SDS Micelle Properties in the Presence of NaCl

In this section we examine the application of the topological algorithm to elucidate the structural evolution of SDS micelles as a function of salt concentration. Fig. 6 depicts the various micelle morphologies extracted from the simulations at different levels of NaCl, where each cluster is coloured as per its shape categorisation derived from the UMMAP package.[47] Our topological algorithm partitions the *extended* UMMAP shape category (orange micelles in Fig. 6) into WLMs and branched micelles, producing the calculated micelle metrics shown

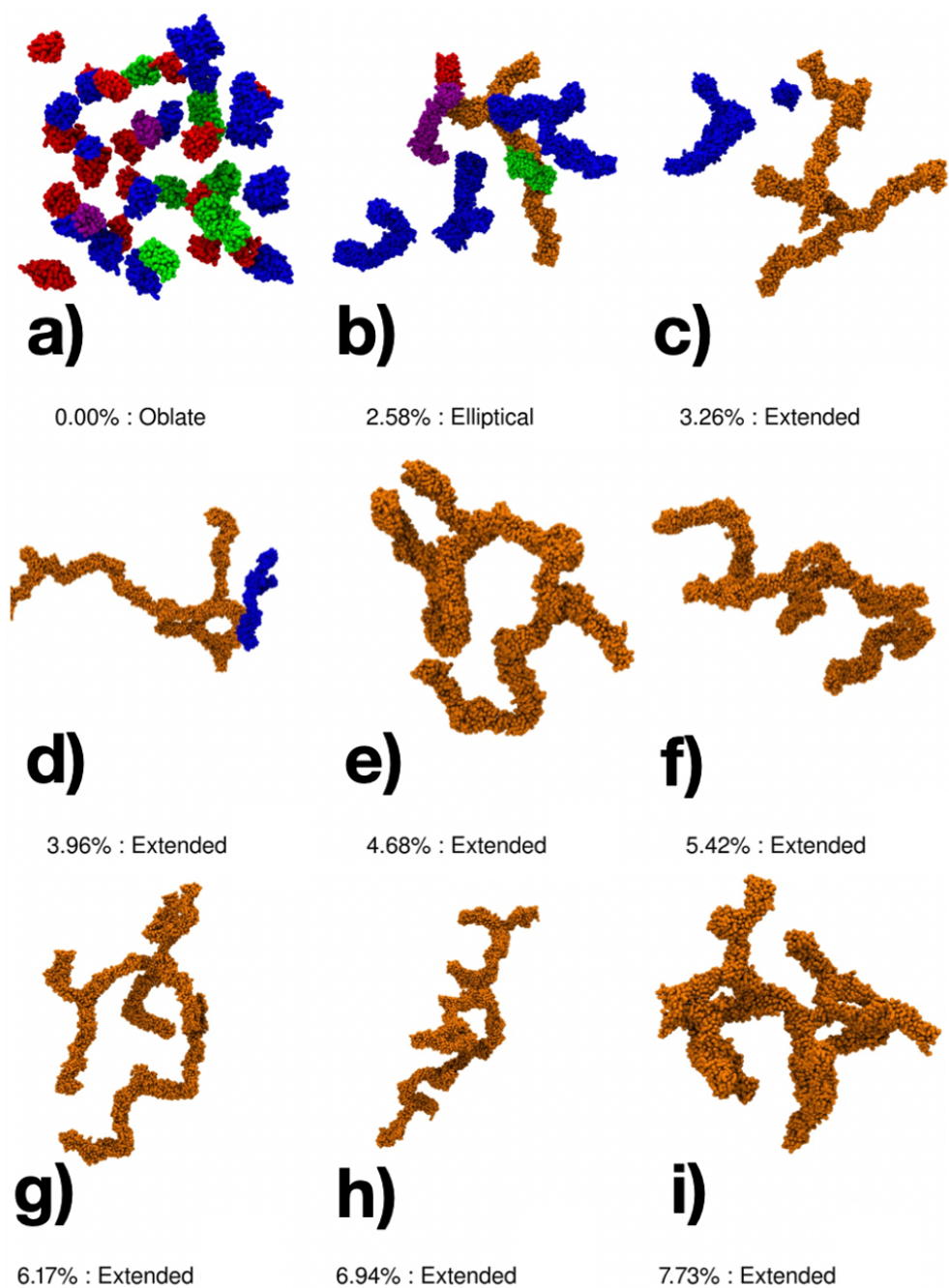

Figure 6: Snapshots of the micelle morphologies of each simulation. The caption of each subplot denotes the salt concentration (w/w %) while the colors denote chemically significant changes in shape.[47]

in Fig. 7.

In the absence of salt (Fig. 6a), the electrostatic interactions between the sulphate head groups produces an unscreened effective repulsive barrier. This inhibits cluster aggregation and results in the formation of small spherical clusters where the surface area is minimised in order to reduce repulsive head group electrostatic interactions.[18] As salt is added to the system, it screens these electrostatic interactions leading to micelle elongation (Fig. 6b) and a reduction in their number (Fig. 7e).

As salt concentration is incremented further, the disruption of the electrostatic repulsion facilitates the transition from rod-like to WLM micelles (Fig. 6b & c). Eventually, it is screened to such an extent that branching node points can develop (Fig. 6c & d) as measured in Fig. 7d. That is to say that branch points where three micelle segments intersect become energetically possible. This signifies the emergence of branched micelles and coincides with further reductions in the number of micelles (Fig. 7e) and expanding micelle backbone networks (Fig. 7b). This trend continues until we end up with one extensive branched micelle which is progressively developing more node points (Fig. 7f).

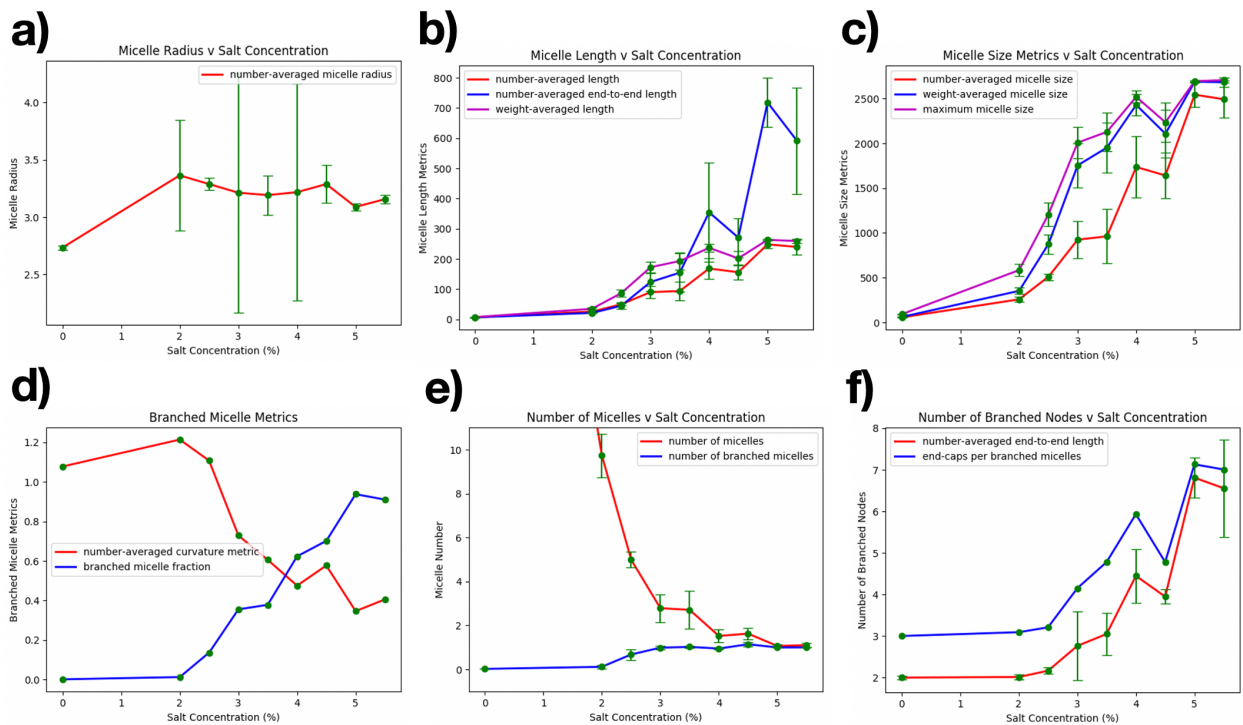

Figure 7: Plots of the calculated micelle structural metrics evaluated for the test case salt curve scan.

Note that we examine the equilibrated values for the structural features and precise information about merging events is neglected in this work (interested readers are directed towards the following references[18,21]).

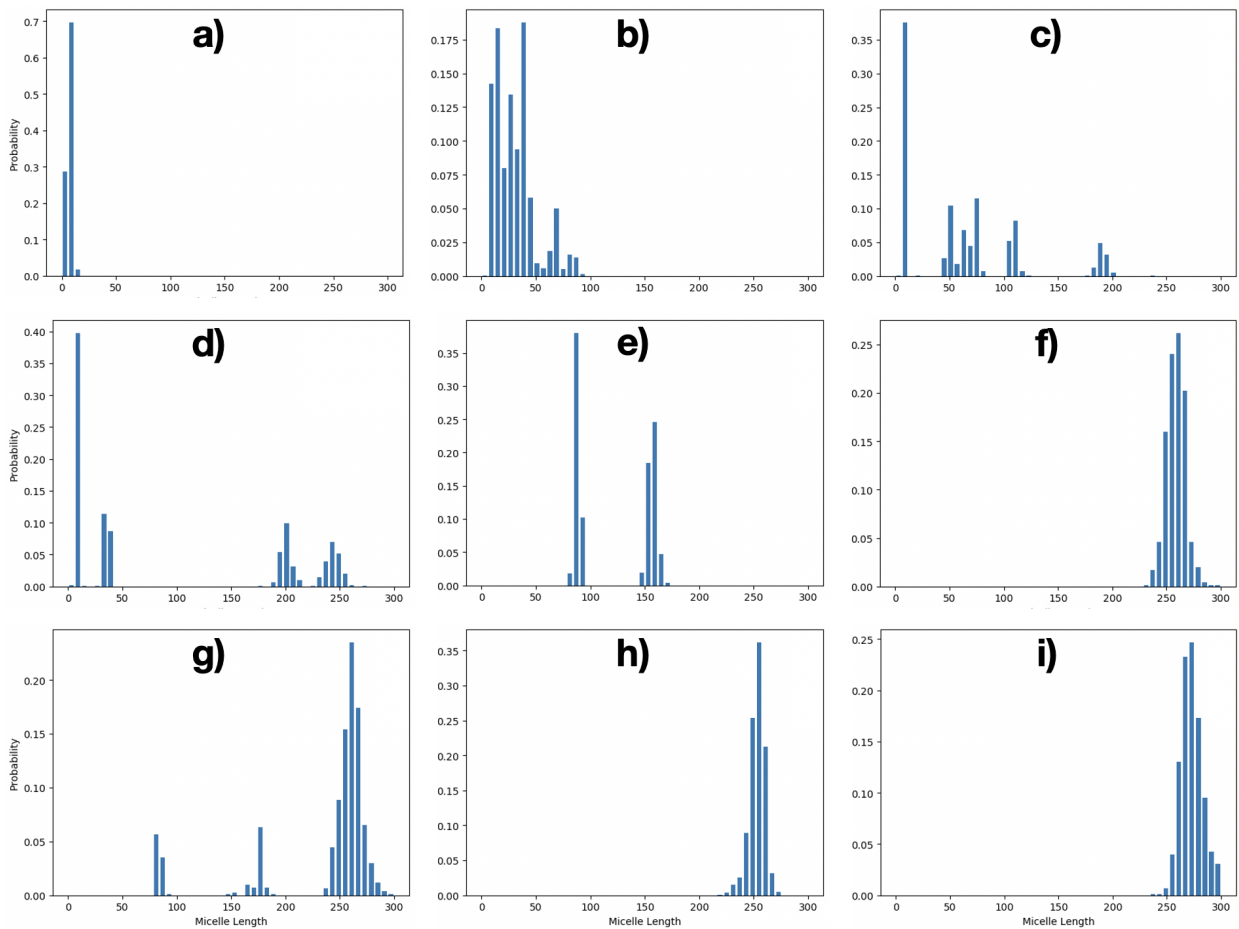

Figure 8: Micelle backbone length probability distributions for a variety of different salt concentrations (w/w %). (a) 0%, b) 2.58%, c) 3.26%, d) 3.96%, e) 4.68%, f) 5.42%, g) 6.17%, h) 6.94% and i) 7.73%.

While the size and length of micelles increases drastically with the addition of salt, the micelle radius remains relatively constant throughout as shown in Fig. 7a. This result is consistent with classical micelle theory[54–57] which dictates that the radius of a rod-like or worm-like micelle is independent of its backbone length. As expected, this characteristic equilibrium length scale was estimated to be a little less than the maximum length of a single SDS molecule[58] (2nm ≈ 3.54 DPD units).

The continuous progression of the micelle morphologies from spherical to wormlike to branched configurations is a gradual as opposed to a discontinuous phase transition. The wide variance in the micelle length distributions displayed in Fig. 8 indicates that a heterogeneous mixture of micelle shapes and sizes is possible at thermodynamic equilibrium. This observation tallies with classic micelle models which theorise monodisperse size distributions for spherical micelles and polydisperse mixtures for elongated structures.[59] It also explains the gentle increase in the fraction of branched micelles displayed in Fig. 7d. Another more subtle sign of this phenomenon can be deduced from the micelle curvature ratio parameter $\alpha$. In this framework, the micelles are worm-like as opposed to branched and therefore this parameter must exceed one by definition. However, this value can be substantially less than unity for branched topologies as evidenced in the plot. It follows that the salt concentration where this parameter equals one can be construed as an approximate location for this phase transition. This corresponds to a transition point between the third and fourth data point in Fig. 7d.

## 5 Discussion

Although micelle length is known to be a key metric associated with structural changes in rods and WLMs [60] there has been no obvious way to measure the length of WLMs. For a cylinder, measuring the axis provides an obvious overall length. Analogously, for WLMs we introduce the PL skeleton as a metric comparable to the axis of the cylinder, while still permitting an easy determination of overall length (by the obvious addition of the lengths of all the sub-segments). This skeleton is similar to the medial axis,[61–66] which plays a prominent role in geometric topology algorithms for many engineering applications. However, the approach here avoids the topological instability problem where small changes in shape cause significant variations in the computed medial axis.[67] The effectiveness of this length value as a metric is corroborated by the chemical simulations reported in the previous section.

Our procedure contains three parameters controlling spectral segmentation which were introduced to avoid observed problem cases of data fitting and whose values were empirically determined. The lower bound on the spectral gap, $\lambda_{min}$ avoids under-fitting while the upper bound, $\lambda_{max}$, and the maximum recursion limit, $k_{max}$ avoid overfitting. In our simulated experiments we found that a balance of these parameters was essential to maintaining good performance. After tuning these parameters we still encountered cases in which spectral segmentation alone failed. An additional repair and merge step was introduced to handle these, which included branch points with high degree and the presence of cycles in the skeleton graph.

The algorithm was designed from a geometric topology perspective with assumptions made about the shape. We assume that the input graph is connected and represents a solid object, with high connectivity in the interior. This assumption is vital to the effectiveness of spectral segmentation as it allows the boundary and branch points to be clearly identified. While our empirically determined parameter values work well in practice under these shape assumptions, we hypothesize that these thresholds reflect basic structural characteristics of the micelles and could be derived directly from a more careful chemical analysis.

Our topological algorithm can be successfully applied to convex micelle geometries, WLMs and branched structures, however there are limitations to its range of applicability. In particular, surfactant bilayer and vesicle micelles were not investigated in this study and further work would be required to extend the mathematical framework such that it could correctly characterise such point cloud configurations.

A key motivation for our algorithm is to accurately measure structural characteristics of WLMs that could be used in conjunction with viscosity models to obtain further insight into their rheological properties. For example, based on polymer physics concepts Boek and co-workers developed a framework, MESOWORM, which models WLMs as a continuous string of thin rods, similar to our derivation of a piecewise linear micelle backbone.[68–72] From this they derived an expression for the zero-shear viscosity, $\eta_0$, of unentangled WLMs.[69]

$$\eta_0 = \frac{\pi}{\sqrt{6}} \frac{c_N k_B T}{N+1} \sqrt{\tau_l \tau}, \tag{1}$$

where $c_N$, $N$, $\tau_l$ and $\tau$ are the number of micelle backbone segments per unit volume, number of micelle backbone segments per micelle, longest relaxation time if the WLMs were unbreakable, and effective relaxation time. The first two of these, $c_N$, $N$ can be calculated at each timestep directly from the application of the algorithm.

The longest relaxation time, $\tau_l$, is given as:

$$\tau_l = \frac{\xi b^2 (N+1)^2}{3\pi^2 k_B T}, \tag{2}$$

Here $b$ is the average micelle segment length which can be extracted from our algorithm. $\xi$ is the average friction on a segment and can be estimated as outlined in the literature.[69]

The final parameter is the effective relaxation time $\tau$. This is a function of $\tau_l$ (Eq. 2) and $\tau_b$, the characteristic breakage timescale, which equals $1/(k_m L)$. Here $L$ is the average micelle length produced from our algorithm and $k_m$ is the micelle breakage rate. Although $k_m$ is not measured by our algorithm, the fact it can measure end-caps allows us to detect their formation, and thus provides an avenue for obtaining this value from the same simulations.

We note that the practical use of the above equations will require careful consideration of finite-size effects and related calibration of the simulation box size. These effects directly impact measurements of extrinsic properties like micelle length and can influence intrinsic properties like micelle backbone segments per unit volume. In addition the growth of the micelles may be inhibited by the finite number of surfactants initialised in the simulation box. These issues will be investigated in more depth in future studies of surfactant rheology.

Finally the algorithm's ability to detect micellar phase transitions (see Section 4) could be useful from a theoretical and experimental perspective. Qualitatively, the onset of branching makes additional stress relaxation modes available,[21] lowering viscosity and indicating that the salt concentration corresponding to a viscosity maximum has been passed. Importantly, the onset of branching cannot be detected experimentally and our model could act as a preliminary screening tool for narrowing down the position of the salt curve maximum, and also help to delineate the bounds of applicability of theoretical viscosity models which cannot be employed on branched WLMs.

# 6 Conclusion

The work presented here tailors techniques from computational geometric topology to address the problem of measuring the length of worm-like micelles (WLMs) resulting in a novel contribution to their analysis. The developed algorithm takes molecular point clouds as input and produce clusters, as determined from the Fiedler vector of the discrete Laplacian, to form piecewise linear (PL) skeletons, and to assign a comprehensive length to each WLM.

The new length metric allows calculation of micelle-curvature, discriminates worms from rods, and extends directly to branched WLMs. In addition, the algorithm provides information on other interesting characteristics of worms such as the number of end-caps and branch-points, along with the micelle cross-sectional radius. Since a key motivation for characterising WLMs is understanding their rheological behaviour we also discuss the theorized connections between the characteristics accessed by our algorithm and viscosity, and show how the algorithm can provide input parameters for well-established mesoscopic WLM viscosity models.

Importantly, the technique can be used to characterise WLMs produced by any particle based simulation technique. Here we demonstrated the algorithm's utility by applying it to DPD simulations of the ion-induced micelle phase transitions of SDS, extracting a series of geometric measures which describe the structural evolution of WLMs as salt concentration is varied.

# 7 Acknowledgements

This work was supported by the STFC Hartree Centre's *Innovation: Return on Research Programme*, funded by the UK Department for Business, Energy & Industrial Strategy. DRS and KG acknowledge, with appreciation, partial funding from NSF grant CCF - 1652218, noting that all statements made here are the responsibility of these authors, not of NSF. TJP and KG, acknowledge, with appreciation, generous funding from IBM Research, under its OCR and SUR programs in award IBM-TJP-6328340. Any statements or errors presented are the responsibility of these authors, not of IBM. RLA and DJB thank Patrick Warren for many useful discussions regarding the formation of worm-like micelles and how their structure relates to liquid properties.

# 8 Appendix: Coarse-Grained DPD Model

In this work we adopt the Dissipative Particle Dynamics (DPD) simulation approach in which particles (referred to as DPD beads hereafter) interact *via* soft repulsions and with local pairwise dissipative and random forces which work together as a thermostat.[42] We do not repeat here details of what is now a standard simulation method. Rather, we point the reader to chapter 17 of the textbook by Frenkel and Smit,[44] and the original DPD literature.[41,42,73] An up-to-date perspective on the DPD methodology has been recently presented by Español and Warren.[43] In the DPD simulations carried out in this article we adopt the sodium dodecyl sulphate (SDS) model developed by by Anderson *et al.*[74,75] This model has been adopted and extended by a number of groups to explore the behaviour of ionic surfactants.[76–79]

In their approach each DPD bead represents between 1-3 "heavy" atoms (namely, oxygen, carbon, sodium and chloride in this study). The coarse-grained SDS molecule is represented by eight beads where one bead is the Na$^+$ counter ion and is not bonded to the rest of the surfactant. Water beads in the model are defined by a mapping number of $N_m = 2$, so that each water bead corresponds on average to two water molecules.[80] Following well-established protocol,[42] the density of water beads is set to $\rho r_c^3 = 3$ in reduced DPD units, where $r_c$ is the above-mentioned cut-off distance. The mapping $\rho N_m v_m \equiv 1$, where $v_m \approx 30\,\text{Å}^3$ is the molecular volume of liquid water, then gives $r_c \approx 5.65\,\text{Å}$ in physical units. Na$^+$ and Cl$^-$ beads are each represented by a single bead and can be considered to be partially hydrated ions in the model. Essentially, these ions are represented as positively and negatively charged water beads in the model. This is a coarse approach to the consideration of charged species but has proven to be effective in exploring the micelle behaviour of anionic surfactants,[9,75] and theories on developing improved representations are only just emerging in the literature.[81] Full details of the model and adopted parameters are presented in the SI.

In this model a standard set of reduced units is adopted in which the beads have unit mass, the system is governed by temperature $k_\mathrm{B} T = 1$ (equivalent to 298 K), and the baseline cut-off distance for the short-range soft pairwise repulsion between solvent beads is set as $r_c = 1$. (Note that the model adopted allows deviations from this for non-solvent beads.)

(5) Biswas, S.; Kumari, P.; Lakhani, P. M.; Ghosh, B. Recent advances in polymeric micelles for anti-cancer drug delivery. *Eur. J. Pharm. Sci* **2016**, *83*, 184 – 202.

(6) Mandal, A.; Bisht, R.; Rupenthal, I. D.; Mitra, A. K. Polymeric micelles for ocular drug delivery: From structural frameworks to recent preclinical studies. *J. Control. Release* **2017**, *248*, 96 – 116.

(7) Cagel, M.; Tesan, F. C.; Bernabeu, E.; Salgueiro, M. J.; Zubillaga, M. B.; Moretton, M. A.; Chiappetta, D. A. Polymeric mixed micelles as nanomedicines: Achievements and perspectives. *Eur. J. Pharm. Biopharm.* **2017**, *113*, 211 – 228.

(8) Skoglund, S.; Lowe, T. A.; Hedberg, J.; Blomberg, E.; Wallinder, I. O.; Wold, S.; Lundin, M. Effect of Laundry Surfactants on Surface Charge and Colloidal Stability of Silver Nanoparticles. *Langmuir* **2013**, *29*, 8882–8891.

(9) Tang, X.; Zou, W.; Koenig, P. H.; McConaughy, S. D.; Weaver, M. R.; Eike, D. M.; Schmidt, M. J.; Larson, R. G. Multiscale Modeling of the Effects of Salt and Perfume Raw Materials on the Rheological Properties of Commercial Threadlike Micellar Solutions. *J. Phys. Chem. B* **2017**, *121*, 2468–2485.

(10) Zheng, R.; Liu, G.; Devlin, M.; Hux, K.; chi Jao, T. Friction Reduction of Lubricant Base Oil by Micelles and Crosslinked Micelles of Block Copolymers. *Tribol. Trans.* **2009**, *53*, 97–107.

(11) Spikes, H. Friction Modifier Additives. *Tribol. Trans.* **2015**, *60*, 5.

(12) Zana, R. *Giant micelles*; Taylor & Francis, 2007.

(13) Padalkar, K. V.; Pal, O. R.; Gaikar, V. G. Rheological characterization of mixtures of cetyl trimethylammonium bromide and sodium butyl benzene sulfonate in aqueous solutions. *Journal of Molecular Liquids* **2012**, *173*, 18–28.
24

# Supporting Information:

# An Efficient Algorithm for Topological Characterisation of Worm-Like and Branched Micelle Structures from Simulations

Breanndan O Conchuir,*,† Kirk Gardner,‡ Kirk E. Jordan,¶ David J. Bray,§
Richard L. Anderson,§ Michael A. Johnston,∥ William C. Swope,⊥ Alex
Harrison,† Donald R. Sheehy,# and Thomas J. Peters*,‡

†IBM Research U.K., The Hartree Centre, Daresbury WA4 4AD, United Kingdom
‡Department of Computer Science & Engineering, University of Connecticut, Storrs, CT
06269, United States
¶IBM T.J. Watson Research, Cambridge, Massachusetts, United States
§The Hartree Centre, STFC Daresbury Laboratory, Warrington, WA4 4AD, United
Kingdom
∥IBM Research Ireland, Dublin, Ireland
⊥IBM Almaden Research Center, San Jose, California, United States
#Department of Computer Science, North Carolina State University, Raleigh, NC 27695,
United States

E-mail: breanndan.conchuir@ibm.co.uk; tpeters@engr.uconn.edu

# Algorithmic Theory

In the topological characterisation procedure we represent a micelle as a graph[S1] associated with a collection of points in $\mathbb{R}^3$. Formally, a graph is an an ordered pair of sets

$$G = (V, E),$$

where $V$ is called the *vertex set*, with *vertices*, $v_i$, and $E$ is called the *edge set* with *edges*, $e_{i,j}$. Figure S1 shows a planar graph, with integers indicating vertices and line segments for edges. We consider undirected graphs, so $e_{i,j} = e_{j,i}$.

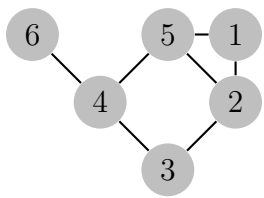

Figure S1: Example of a graph with vertices labeled 1-6.

The *degree* $d_i$ of a vertex $v_i$ is the number of edges in $E$ containing $v_i$. The *degree matrix* of a graph with $n$ vertices is the $n \times n$ diagonal matrix $D$ such that

$$D_{ij} = \begin{cases} d_i & \text{if } i = j \\ 0 & \text{otherwise.} \end{cases}$$

The *adjacency matrix* of $G$ is the $n \times n$ symmetric matrix $A$ with entries

$$A_{ij} = \begin{cases} 1 & \text{if } e_{i,j} \text{ is an edge,} \\ 0 & \text{otherwise.} \end{cases}$$

The *graph Laplacian*[S1–S4] of a graph $G$ is the $n \times n$ matrix $L = D - A$.

In Figure S1 the vertex labeled with 1 has 2 edges and the degree of vertex 1 is 2, for

$$D = \begin{bmatrix} 0 & 1 & 0 & 0 & 1 & 0 \\ 1 & 0 & 1 & 0 & 1 & 0 \\ 0 & 1 & 0 & 1 & 0 & 0 \\ 0 & 0 & 1 & 0 & 1 & 1 \\ 1 & 1 & 0 & 1 & 0 & 0 \\ 0 & 0 & 0 & 1 & 0 & 0 \end{bmatrix} \quad A = \begin{bmatrix} 2 & 0 & 0 & 0 & 0 & 0 \\ 0 & 3 & 0 & 0 & 0 & 0 \\ 0 & 0 & 2 & 0 & 0 & 0 \\ 0 & 0 & 0 & 3 & 0 & 0 \\ 0 & 0 & 0 & 0 & 3 & 0 \\ 0 & 0 & 0 & 0 & 0 & 1 \end{bmatrix} \quad L = \begin{bmatrix} 2 & -1 & 0 & 0 & -1 & 0 \\ -1 & 3 & -1 & 0 & -1 & 0 \\ 0 & -1 & 2 & -1 & 0 & 0 \\ 0 & 0 & -1 & 3 & -1 & -1 \\ -1 & -1 & 0 & -1 & 3 & 0 \\ 0 & 0 & 0 & -1 & 0 & 1 \end{bmatrix}.$$

A standard refinement of the Laplacian defined above which is often used to compare matrices that are of varying size is the normalized symmetric Laplacian,[S5] denoted as $L^{sym}$. The adjustments are given by entries

$$L^{sym}_{ij} = \begin{cases} 1 & \text{if } i = j \text{ and } d_i \neq 0, \\ -\frac{1}{\sqrt{d_i d_j}} & \text{if } v_i, v_j \text{ is an edge} \\ 0 & \text{otherwise.} \end{cases}$$

Throughout, we will denote the normalized laplacian $L^{sym}$ by $L$.

Let $\lambda_0 \leq \lambda_1 \leq \ldots \leq \lambda_{n-1}$ denote the *eigenvalues* of $L$ and $w_i \in \mathbb{R}^n$ denote the *eigenvector* corresponding to $\lambda_i$: that is, $Lv_i = \lambda_i w_i$. For a connected graph $\lambda_0 = 0$ and $\lambda_1$ is the *spectral gap* which gives a measure of the connectedness of the graph. The eigenvector corresponding to the smallest non-zero eigenvalue, $w_1$ if the graph is connected, is known as the *Fiedler vector*.[S2,S6,S7] We will assume $G$ is connected and denote $f = w_1$ and $\lambda = \lambda_1$ to simplify notation.

Vertices will be denoted by $v_i$ for $i = 1, \ldots, n$. Entries of $f$ are denoted $f_i$, for $i = 1, \ldots, n$. The graph $G = (V, E)$ can be split into subgraphs $G_1$ and $G_2$ by partitioning $V$ into

$$V_1 = \{v_i \in V \mid f_i < 0\}, \quad V_2 = \{v_i \in V \mid f_i \geq 0\},$$

and taking $G_1$ and $G_2$ as induced subgraphs of $G$.

Formally, a *graph partition* is the reduction of a graph $G = (V, E)$ into a collection of smaller graphs with $H = (U, L)$ formed by partitioning its vertices $V$ into mutually exclusive, non-empty, vertex sets $U = V_i, \ldots, V_m$. That is, each vertex of $H$ is a subgraph $H_i = (V_i, E_i)$ such that $V_i \cap V_j = \emptyset$ for all $i \neq j$ and $\bigcup_{i=1,\ldots,n} V_i = V$ with edges $E_i$ consisting of edges in $G$ with both ends in $V_i$. The edges $L$ of the graph partition are given by collections of edges in the initial graph that cross between groups and will be referred to as *links*.

A single partition of the vertices a connected graph $G = (V, E)$ into two groups $V_1$ and $V_2$ gives a graph partition $H = (\{G_1, G_2\}, \{l\})$. The link $l$ associating $G_1$ and $G_2$ is itself a subgraph of $G$ given by the collection of edges

$$E_l = \{e = \{v_i, v_j\} \in E \mid f_i < 0 \text{ and } f_j \geq 0\}$$

and the vertex set

$$V_l = \bigcup_{e \in E_l} e.$$

The process of repeatedly partitioning subgraphs split by this process will be referred to as *recursive spectral segmentation*. Empirically, we will introduce a bound on the spectral gap as a stopping condition for the recursion.

## Molecular Beading

A total of three molecules, SDS, NaCl and water, are simulated in this study. The atomic to DPD bead mapping is outlined in Table S1, while each water bead consists of two water molecules. Similarly, a DPD NaCl molecule is composed of two unbound DPD beads: Na and Cl. Both of these ions are denoted to be hydrated and consist of two atomic water molecules along with the corresponding ionic atom.

The construction of the DPD SDS molecule is illustrated in Fig. S1. This linear molecule has a radial bond $U(r_{i,j})$ between each neighbouring pair of beads $i$ and $j$, as per Eq. 7. Here $\kappa$ is set to 150 and $r_0$ is defined by the equation $r_0 = (n_i + n_j) - 0.01$ where $n_i$ is the number of "heavy" (non-hydrogen) atoms per bead. The equilibrium bond angle $U(\theta_{ijk})$ is set to 180 degrees for all angular bonds, along with a stiffness parameter $\kappa$ set to 5.

Table S1: Table mapping each DPD bead to its constituent atomic components.

| DPD Bead | Atoms | Charge |
|---|---|---|
| $H_2O$ | $H_4O_2$ | 0 |
| $Na$ | $NaH_4O_2$ | 1 |
| $Cl$ | $ClH_4O_2$ | -1 |
| $CH_2SO_4$ | $CH_2SO_4$ | -1 |
| $CH_3$ | $CH_3$ | 0 |
| $CH_2CH_2$ | $C_2H_4$ | 0 |

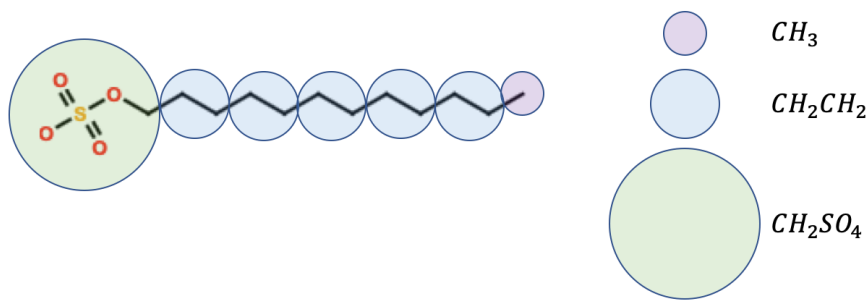

Figure S2: Schematic of the beading structure of SDS.

## Force Field Parameters

The force-field parameters employed in this study are displayed in Table S2.[S8]

## Mathematical Derivations of Simulation Observables

The simulation observables which we computed in the study are as follows:

- The number-averaged micellar cross-sectional radius $r_{cs}$.

- The number-averaged micelle backbone length $l_b$.

Table S2: Table of DPD interbead interaction parameters as per Eq. 2.

| Bead $i$ | Bead $j$ | $A_{ij}$ | $r_c$ |
|---|---|---|---|
| $Cl$ | $Cl$ | 25.00 | 1.00 |
| $Cl$ | $Na$ | 25.00 | 1.00 |
| $Cl$ | $H_2O$ | 25.00 | 1.00 |
| $Cl$ | $CH_3$ | 45.00 | 0.98 |
| $Cl$ | $CH_2SO_4$ | 25.00 | 1.00 |
| $Cl$ | $CH_2CH_2$ | 45.00 | 1.04 |
| $Na$ | $Na$ | 25.00 | 1.00 |
| $Na$ | $H_2O$ | 25.00 | 1.00 |
| $Na$ | $CH_3$ | 45.00 | 0.98 |
| $Na$ | $CH_2SO_4$ | 17.90 | 1.12 |
| $Na$ | $CH_2CH_2$ | 45.00 | 1.04 |
| $H_2O$ | $H_2O$ | 25.00 | 1.00 |
| $H_2O$ | $CH_3$ | 45.00 | 0.98 |
| $H_2O$ | $CH_2SO_4$ | 17.90 | 1.12 |
| $H_2O$ | $CH_2CH_2$ | 45.00 | 1.04 |
| $CH_3$ | $CH_3$ | 24.00 | 0.95 |
| $CH_3$ | $CH_2SO_4$ | 28.50 | 1.09 |
| $CH_3$ | $CH_2CH_2$ | 23.00 | 1.01 |
| $CH_2SO_4$ | $CH_2SO_4$ | 13.30 | 1.23 |
| $CH_2SO_4$ | $CH_2CH_2$ | 28.50 | 1.15 |
| $CH_2CH_2$ | $CH_2CH_2$ | 22.00 | 1.07 |

- The weight-averaged micelle backbone length $l_{wb}$.

- The number-averaged micelle end-to-end length $l_c$.

- The number-averaged micelle size $N_s$.

- The weight-averaged micelle size $N_{ws}$.

- The maximum micelle size $N_{max}$.

- The number of micelles $N_m$.

- The number of branched micelles $N_{bm}$.

- The micelle curvature ratio $\alpha$.

- The number-averaged end-caps per micelle $N_{end}$.

## Structural Observables

The following structural observables were calculated directly from the simulation itself:

- The maximum micelle size $N_{max}$.

- The number of micelles $N_m$.

- The number of branched micelles $N_{bm}$.

The remaining structural observables were derived from direct observables (parameters which the simulation measured at each time step).

$$
\begin{aligned}
r_{cs} &= \frac{r^1_{mcs}}{r^0_{mcs}}, \\
l_{nb} &= \frac{l^1_{mb}}{l^0_{mb}}, \\
l_{wb} &= \frac{l^2_{mb}}{l^1_{mb}}, \\
l_e &= \frac{l^1_{me}}{l^0_{me}}, \\
N_{ns} &= \frac{N^1_{ms}}{N^0_{ms}}, \\
N_{ws} &= \frac{N^2_{ms}}{N^1_{ms}}, \\
\alpha &= \frac{l_{nb}}{l_c}, \\
N_{end} &= \frac{N^1_{end}}{N^0_{ms}}.
\end{aligned}
\quad (1)
$$

Here $r^i_{mcs}, l^i_{mb}, l^i_{me}, N^i_{ms}, N^i_{end}$ signifies the $i$th moment of the micelle cross-sectional radius, micelle backbone length, micelle end-to-end length, micelle size, micelle end-cap distributions, respectively.

## Simulation Equilibration Protocol

A sliding window weighted linear-least squares (WLLS) procedure was applied to determine the equilibration time (i.e. time after which a steady state value is achieved) of each measured time series observable.[S9] The equilibration point of the time series which converges slowest to equilibrium is defined as the global equilibration point. The set of simulation frames from this point until the end of the simulation was defined as the range of equilibrated frames, and it was from this that the mean, correlation time and standard error of the mean of each time series observable was calculated. Each simulation ran for 8 million time steps to ensure that the number of equilibrated frames exceeded at least six of the longest observable autocorrelation timescales, and reliable statistics could be gathered.[S10] A detailed explanation of this protocol is beyond the scope of this paper and may be found in the following reference.[S9]